\documentstyle[aps,12pt]{revtex}
\begin{document}
\title{Test of a simple and flexible S$_8$ model molecule in   
$\alpha $-S$_8$ crystals}
\author{C. Pastorino and Z. Gamba}
\address{Department of Physics, Comisi\'on Nacional de 
Energ\'\i a At\'omica, \\
Av. Libertador 8250, (1429) Buenos Aires, Argentina. e-mails:\\
clopasto@cnea.gov.ar 
and gamba@cnea.gov.ar\\}

\maketitle

\begin{abstract}

 $\alpha $-S$_8$ is the most stable crystalline form,  
 at ambient pressure and temperature (STP), of elemental sulfur. In this 
paper we analyze the zero pressure low temperature part
of the phase diagram of this 
crystal, in order to test a simple and flexible model molecule.
The calculations consist in a series of molecular dynamics (MD) 
simulations, performed in the constant pressure- constant temperature 
ensemble. Our calculations show that this model, that gives good 
results for three crystalline phases at STP
 and T$_{\sim }^{>}$300K, fails at low temperatures, 
predicting a structural phase transition at 200K where 
there should be none.\\

\end{abstract}

{\bf Introduction:}

Elemental sulfur is of relevance in geophysics, astrophysics, 
material sciences and massively used in industry  
\cite{steudel,meyer1}. Recently, a high pressure superconductor 
transition, with T$_c$=17K, has been found \cite{highP-S}.

Elemental sulfur is widely known by the unusual characteristics 
of its liquid phase. Numerous works have been dedicated to the
study of its liquid-liquid ($\lambda $) phase transition, due 
to a large variation of
its molecular composition. The $\lambda $ transition is
 of relevance in numerous
fields, volcanology is an example \cite{volcan1,volcan2,volcan3}. The 
liquid phase is also of enormous theoretical interest, with studies 
ranging from models to describe 
 the $\lambda $ transition \cite{trans-liq1,trans-liq2,trans-liq3},
 to molecular dynamics (MD) simulations of the 
liquid \cite{mdliq1,mdliq2,mdliq3} and to calculations of the
stability and structure of isolated S$_n$ molecules 
\cite{mo1,mo2,mo3,mo4}. 
 
In spite of the numerous experimental and theoretical
studies on the liquid phase, the phase diagram and properties
of the crystalline phases of elemental sulfur are far from 
well known \cite{steudel,meyer1}. A simple inter- and 
intramolecular potential, able to reproduce the structural
and dynamical properties of the crystalline phases, should
be very useful to understand the complex experimental data 
of sulfur crystals.

The most stable sulfur allotrope at ambient temperature and 
pressure (STP) is S$_8$, a cyclic crown-shaped  molecule, 
stable in solid, liquid and gas phases. Three crystalline 
phases have been clearly identified at zero pressure and 
T$_{\sim }^{<}$400K, being orthorhombic 
$\alpha $-S$_8$ the most stable crystalline form at STP. 
 Fig. 1a shows its "crankshaft" structure
with four molecules in the primitive unit cell. The space group is D$%
_{2h}^{24}$ (Fddd), with 16 molecules in this non-primitive 
orthorhombic cell. If 
$\alpha $-S$_8$ is slowly heated, it shows a phase transition to monoclinic 
$\beta $-S$_8$ at 369K, which melts at 393K.\cite{steudel}. Nevertheless, 
most $\alpha $-S$_8$ crystals do not convert to $\beta $-S$_8$
 \cite{beta2}, 
they melt, instead, at 385.8K \cite{steudel,meyer1}.
% $\beta $-S$_8$ is a monoclinic crystal, with six molecules in the 
% primitive unit cell, two of them orientationally disordered
% \cite{beta1,beta2}. A third crystalline allotrope has been observed: 
% $\gamma $-S$_8$ (Fig. 1c), that can be obtained from solutions of 
% S$_8$ or from its melt \cite{steudel,meyer1}. The space group 
% is C$_{2h}^{4}$ (P2/c), with four molecules in the pseudo-hexagonal 
% closed-packed unit cell \cite{gam1}. The density of this allotrope, 
% at STP, is 5.8 \% higher than that of $\alpha $-S$_8$.

Experimentally, the samples of pure $\alpha $-S$_8$ can be 
easily identified because these crystals are greenish-yellow at 
293K and turn white below 193K. Commercial sulfur, instead,
 contains small amounts of S$_7$ and conserves the yellow 
color at low temperatures \cite {steudel}. The change of color 
is assumed to be due to a shift of the electronic band gap
between valence and conduction bands  \cite{steudel,meyer1}.

 The firsts molecular dynamics (MD) simulations of 
$\alpha $-S$_8$ crystals were performed in the constant volume-
constant energy ensemble,
at normal pressure and at several temperatures \cite{cardini1}, with
the only aim to study the Raman and ir. intramolecular frequencies
in a crystalline enviroment.
 Numerous works \cite{alfafreq} have also been 
dedicated to the measurement and/or calculation, by lattice sums 
and in different approximations, of the 
{\bf k}=0 vibrational modes.

The first attempt to reproduce the zero pressure phase diagram of
$\alpha $-, $\beta $- and $\gamma $-S$_8$ crystals has been 
recently performed \cite{paper1}, using a simple and 
flexible model molecule, based in that proposed by Venutti et al.
 \cite{cardini1}. The model molecule reproduces the STP structures the
three crystals, the measured heat of sublimation of
 $\alpha $-S$_8$, and the orientationally disordered high 
temperature plastic phase of $\beta $-S$_8$ crystals. 

In this paper we study the zero pressure and low temperature 
phase diagram of $\alpha $-S$_8$ crystals,
as given by the simple and flexible model molecule of ref. 
\cite{cardini1}. The calculations
consist of a series of constant pressure - constant temperature 
classical MD simulations, at several temperatures and zero pressure. This
MD algorithm is useful to locate structural phase transitions,
or in the search of the most stable structure for a given set
of thermodynamic parameters P and T.

Unfortunately, the model molecule of ref. \cite{paper1} 
fails to reproduce
the measured low temperature structure of  $\alpha $-S$_8$ crystals:
 a structural phase transition is found at 200K, were there
should be none. Here we give a detailed account of the 
calculations performed at low temperatures on $\alpha $-S$_8$ crystals,
of a disordered sample,  
and of the reasons for the failure of the proposed model.\\

{\bf The inter- and intramolecular potential model:}

 The main problem found in this type of calculations 
is the inter- and intramolecular potential model of this 
flexible molecule. The model of our simulations is a slight
modification of that proposed in the constant volume MD 
simulations of $\alpha $-S$_8$ crystals, that succesfully 
reproduced the available experimental data on the crystalline 
intramolecular frequencies \cite{cardini1}.
 The intermolecular potential model of ref. \cite{cardini1} was 
of the Lennard-Jones (LJ) atom-atom type. In our constant pressure-
 constant temperature MD calculations, these LJ 
parameters \cite{cardini1} were slightly changed to improve 
the fit of the calculated STP crystalline structure and 
 configurational energy of $\alpha $-S$_8$ to the  
experimental data. The parameters of our LJ
potential model, for non-bonded S-S interactions are: 
$\varepsilon $= 1.70kJ/mol and $\sigma $=3.39 \AA . The cut-off
radius of atom-atom interactions is 12\AA\  and correction
terms to the configurational energy and pressure, due to this
finite value, are taken into account in the usual way: by
integrating the contribution of an uniform distribution of atoms.

A similar procedure is followed for the intramolecular potentials 
of the bending and torsional angles, that are
based on those of ref. \cite{cardini1}. Our flexible model 
molecule takes into account all the low frequency intramolecular
 modes that mix with lattice modes and can
therefore be relevant to the onset of structural changes as
a function of T and P values.
Since the streching modes (from 400 to 500cm$^{-1}$ ) are always 
well above in energies than the rest of the vibrational 
modes ($<$ 250cm$^{-1}$) and of the lattice modes 
(between 10 and $\sim $100cm$^{-1}$), the S-S bond length of our 
model molecule is held constant at 2.0601\AA. The bending and torsional
intramolecular motions (from 150 to 250cm$^{-1}$), instead,
 are taken into account and their mix with lattice modes is 
allowed in a  straigthforward way.

 The intramolecular potential for bending angles S-S-S is
 taken harmonic
\[
V(\beta )=\frac 12C_\beta (\beta -\beta _0)^2, 
\]
with a force constant of C$_\beta $=25700Kk$_B/$rad$^2$ and $\beta _0$= 108deg$%
. $ The intramolecular potential for torsion ($\tau $) angles is a double
well:
\[
V(\tau )=A_\tau +B_\tau \cos (\tau )+C_\tau \cos ^2(\tau )+D_\tau \cos
^3(\tau )\text{,} 
\]
with $A_\tau $=57.19 k$_B$, $B_\tau $=738.41 k$_B$, $C_\tau $=2297.90 k$_B$ 
and $D_\tau $=557.25 k$_B$. These parameters describe a double well with minima at 
$\tau $=${^+_-}$ 98.8deg., and a height barrier of about 12kJ/mol.\\

{\bf Calculations:}

The structural and dynamical properties of $\alpha $ -S$_8$,
are studied in the (N,P,T) ensemble, $via$ a series of classical constant
pressure - constant temperature MD\ simulations. The standard 
MD algorithm allows volume and shape fluctuations of the MD sample 
in order to balance the applied isotropic external pressure with 
the internal stresses \cite{algor2}, the temperature control of the sample
follows the approach of Nos\'e \cite{algor3,algor4}.
The equations of motion of these flexible molecules are
integrated using the Verlet algorithm for the atomic displacements and the 
Shake algorithm for the constant bond lengths constraints on each 
molecule\cite{algor1,tildesley}. The final MD\ algorithm is identical 
to that used in a study of black Newton films \cite{bubbles1}.

The starting point of our simulations is the experimental structure 
of $\alpha $-S$_8$ at STP \cite{alfa1}. The sample was first
equilibrated  with a constant volume algorithm on a trajectory 
of 30000 time steps (of 0.01ps.). Afterwards, the runs in the 
(N,P,T) ensemble were performed by decreasing and increasing
the temperature (up to 450K) in steps of 25 or 50K. At each point of the
phase diagram the sample is equilibrated for 20000 to 30000 time steps and 
measured in the following 10ps. Near the phase transitions the equilibration
times were increased several times.

The sample of $\alpha $-S$_8$ crystals consisted of 3x3x2
orthorhombic cells (288 molecules). Some points of the phase
diagram were recalculated with a sample of 4x4x2 cells (512 molecules).\\

{\bf Results:}

{\bf a) The }$\alpha ${\bf \ -phase:}

$\alpha $-S$_8$ crystallizes, at STP, in the orthorhombic space group F$_d$
(D$_{2h}^{24}$) with 16 molecules in a face-centered cell, the lattice 
parameters at 300K are a=10.4646\AA, b=12.8660\AA\ and c=24.4860\AA
 \cite{alfa1}.  The
primitive cell contains 4 molecules at C$_2$ sites. The molecular rings lie
parallel to the crystallographic $c$ axis and are alternately oriented
parallel to (110) and (1\=10) crystalline planes. Fig.1a includes part
of our sample at 300K and shows the structure in the [110] direction. At
300K our calculated lattice parameters are a=10.34(10)\AA, 
b=13.20(3)\AA\ and c=24.20(5)\AA, the
largest difference is found in the value of b axis, 2 \% higher than its
experimental value. The molecules location and  
orientation follow the experimental data \cite{alfa1}.  At 300K 
the calculated value of configurational energy is 
-103.5(2)kJ/mol, 
 from which the heat of sublimation can be estimated as 101.0kJ/mol
and compares well with a measured value of 101.8(2)kJ/mol \cite{hsubalfa}. 

Figs. 2a and 2b show the calculated vibrational density of states 
of the isolated molecule and that of $\alpha $-S$_8$ at 300K.
They reproduce the experimental pattern, although 
the intramolecular modes are calculated about 10\% higher than
the experimental ones.\\

Figs. 3a and 3b show the calculated configurational energy and 
volume per molecule as a function of temperature, for both 
samples of $\alpha $-S$_8$.
Unfortunately, our orthorhombic $\alpha $-S$_8$ sample
 of 288 molecules turns 
out to be unstable for T$\leq $200K and distorts to a monoclinic cell. 
The final molecular array closely resembles that 
of $\alpha $-S$_8$ (Fig. 1b). Fig3c shows the calculated
angles of the unit cell as a  function of temperature. 
 This result was later on confirmed with a 
larger sample of 512 molecules (4x4x2 orthorhombic cells). Below 200K this 
large sample do not distort to monoclinic, but nevertheless its 
configurational energy and volume per molecule are higher 
than those for the sample of 288 molecules (Figs. 3a and 3b).
 The fluctuations of energy 
and volume, in the sample of 512 molecules,
 are also larger at T$\leq $200 than those at 200K.

 The coincidence of the experimental change of color at 193K
and our calculated transition at 200K, made us review 
the experimental data on this phase.
 The usual reference quoted for the 
lattice parameters values, as a function of temperature, is 
ref.\cite{alfa2}. Their reported values were determined by X-ray powder 
diffractrometry, on different samples and by measuring only three lines
at 8 temperatures between 100 and 300K.
Our calculated powder diffraction patterns, not included here,
are quite similar for the orthorhombic and monoclinic samples
and no conclusive answer can be obtained from these data.
 Nevertheless, there is a careful single crystal determination 
of the structure of $\alpha $-S$_8$ at 300 and 100K \cite{alfa3},
where the crystalline charge distribution is measured. Both
measured structures belong to the same crystalline space group
and the molecular arrangement is identical. This study 
clearly disregard any possible structural phase transition and shows
that the simple model molecule, in spite of the amount 
of experimental data that it is able to reproduce for 
T$\geq $200K, should still be improved \cite{paper1}.\\

{\bf b) Phase transitions and the liquid phase:}

The experimental evidence shows that the phase transitions of 
 S$_8$ crystals are obtained when defects are present, if they 
are not, metastable states can be maintained for 
days \cite{steudel,meyer1,beta2}. This evidence is valid for
the $\alpha -\beta $ transition, and also for the solid-liquid 
phase transition \cite{meyer1}. In the last case, the disorder is
generated because the S$_8$ molecules start to dissociate 
when the temperature is increased.

In our simulations, all calculated ordered samples of $\alpha $-S$_8$
 crystals melt at T$\geq $450K. That the disorder is essential 
to promote the solid-liquid phase transition can be checked by 
simulating a disordered sample of S$_8$ molecules. A cubic
sample of 216 disordered S$_8$ molecules was studied in the range
350-450K, following the same procedure as for the other samples. In
this case the MD algorithm allowed changes in the volume of
the MD box, but not in its shape. Fig. 4 shows the calculated 
discontinuity in the configurational energy and volume per 
molecule at 400K, near that measured for 
 $\alpha $-S$_8$ (386K) and  $\beta $-S$_8$ (393K) 
crystals. For T$\geq $400K the sample is liquid, as was determined 
by its diffusion constant. These simulations should be valid 
for temperatures above and near the transition, when the 
fraction of broken molecules is small, and show that the model
molecule is useful in this range of temperatures.\\

The calculated distances between firsts neighbor atoms compares well 
with those determmined from the pair distribution function g$_2$(r) 
obtained by neutron measurements \cite{egelstaff}. Fig. 4a includes 
our calculated bonded and non-bonded atom-atom distances of 
crystalline $\alpha $-S$_8$ at 300K, Fig. 4b corresponds to 
our disordered 'liquid' sample at 425K. The 3$^{th}$ 
and 5$^{th}$ peaks at 3.9\AA\   and 5.4\AA , which are not 
found in the liquid phase, correspond to non-bonded distances.
Their disappearance, observed in the liquid phase, cannot be unequivocally
 attributed to broken bonds \cite{egelstaff}. \\

{\bf Conclusions:}
 
In this paper we study the zero pressure low temperature range
of the phase diagram of S$_8$ crystals, as given by a simple 
and flexible model molecule. 
The molecule model, a simplified and slight modification of that
 proposed in ref. \cite{cardini1}, is extremely simple and 
gives good account of many experimental facts for 
T$\geq $200K, including the experimental data on the structure and 
dynamics of $\alpha $-, $\beta $- and $\gamma $-S$_8$ crystals 
around STP and the orientationally disordered plastic phase 
of $\beta $-S$_8$ crystals \cite{paper1}. It also
reproduces the solid-liquid phase transition, near the
experimental temperature, but only for a disordered sample of
S$_8$ molecules \cite{paper1}. The last calculation is only of theoretical 
interest, and we do not intended, here, to simulate the real
liquid phase of elemental sulfur. Nevertheless, the pair 
correlation function calculated in this theoretical 'liquid'
shows that the 3$^{th}$ and 5$^{th}$ neighbours identified in solid
phases correspond to non-bonded atoms, and that their 
disappearance at high temperatures can be expected even if the 
molecules do not dissociate in short segments.\\

 In spite of all the properties of S$_8$ crystals
that the simple and flexible model molecule reproduces,
 below 200K this model is unable to calculate the experimental
 orthorhombic structure of $\alpha $-S$_8$. Although no structural 
phase transitions are measured for $\alpha $-S$_8$ crystals,
 our sample of 288 molecules is unstable and transforms to a 
monoclinic structure.\\

{\bf Acknowledgement:}
The authors thank CONICET for the grant PIP 0859/98.\\

\newpage

{\bf Figures:}

Fig. 1: Calculated stable structures, as given by this model molecule:
 a) orthorhombic $\alpha $-S$_8$ at 
300K, b) monoclinic $\alpha $'-S$_8$ at 150K.\\

Fig. 2: Vibrational density of states: a) Isolated molecule and b)
 $\alpha $-S$_8$ crystal, both at 300K.\\

Fig. 3: Calculated (a) configurational energy and (b) volume per 
molecule of $\alpha $-S$_8$ (large circles: sample of 288 molecules
and small circles: sample of 512 molecules) and  a disordered cubic
 sample (triangles: increasing T, squares: decreasing T) 
 as a function of temperature. For T$\geq $400K, the
disordered cubic sample is liquid. (c) Unit cell angles of 
the $\alpha $-S$_8$ sample of 288 molecules. The lines are 
a guide to the eyes.\\

Fig. 4: Pair correlation functions: a) $\alpha $-S$_8$ at 300K,
b) our theoretical 'liquid' sample of S$_8$ molecules at 425K.
 Full lines: total atom-atom g$_2$(r) functions,
 dotted lines: excluding intramolecular distances.\\
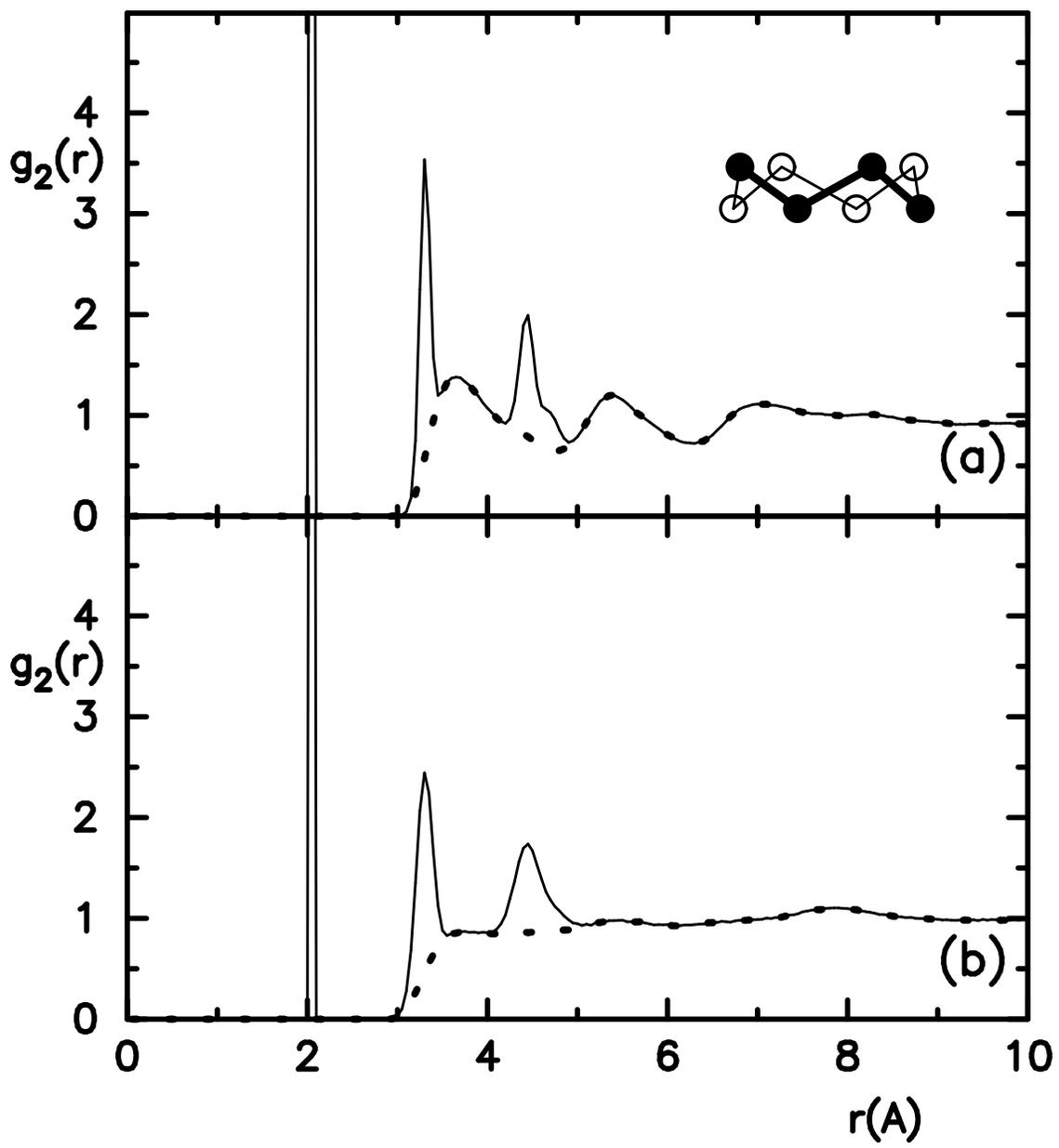
\end{document}